\def\fun#1#2{\lower3.6pt\vbox{\baselineskip0pt\lineskip.9pt
  \ialign{$\mathsurround=0pt#1\hfil##\hfil$\crcr#2\crcr\sim\crcr}}}
\relax
\documentstyle[12pt]{article}
\advance\textheight by \topskip
\begin{document}
%\bf
\begin{flushright}
SU-ITP-93-37\\
UG-8/93\\
hep-th/9401025
\end{flushright}
\vspace{-0.2cm}
\begin{center}
{\large\bf EXACT DUALITY\\
\vskip .8 cm
IN STRING EFFECTIVE ACTION }\\
\vskip 1.7 cm
{\bf Eric Bergshoeff \footnote{ E-mail:
bergshoe@th.rug.nl}\,  and\, Ingeborg Entrop}
\footnote{E-mail: entrop@th.rug.nl}
\vskip 0.05cm
Institute for Theoretical Physics, University of Groningen\\
Nijenborgh 4, 9747 AG Groningen, The Netherlands
\vskip .5truecm
{\bf Renata Kallosh} \footnote {  E-mail:
kallosh@physics.stanford.edu}
 \vskip 0.05cm
Physics Department, Stanford University, Stanford   CA 94305\\
\vskip 0.7 cm
\end{center}
\vskip 1.5 cm
\centerline{\bf ABSTRACT}
\begin{quotation}
We formulate  sigma-model duality transformations in terms of spin
connection.
This allows to investigate the symmetry of the string action
including higher
order   $\alpha'$ corrections. An important  feature of the new
duality
transformations is a simple homogeneous
transformation  rule of the spin connection (with torsion) and
specifically
adjusted
transformation of the Yang-Mills field.

We have found that under certain conditions  this duality is a
symmetry of the full effective string action in the target space,
free of
$\alpha'$
corrections. We demonstrate how the exact duality generates new
fundamental
string
solutions from supersymmetric string waves.

\end{quotation}

%\normalsize
\newpage
\section{Introduction}

In recent years an active field of research has been
 the study of modified Einstein-Maxwell equations.
 The modifications that have been considered
 include additional scalar or an\-ti\-sym\-me\-tric tensor
 fields (called dilatons and axions)
or modifications in which the
 electromagnetic field is replaced by non-Abelian Yang-Mills
 fields. These modified Einstein-Maxwell theories admit new
 solutions whose consistency crucially depends on the presence of
 the new fields and/or on the non-Abelian nature of the Yang-Mills
 fields. For examples of such new solutions see
 the review articles \cite{Ca1,Ho2,Se1} and references
therein.

One motivation for studying the above-mentioned modifications
 to Einstein-Maxwell theory is that they arise in string theory.
 The zero slope limit $\alpha'\to 0$ of string theory
corresponds to a modified
 Einstein-Maxwell theory of the type discussed above.
 The complete effective
 action also includes contributions which are of higher order in
 the Riemann tensor
 and the Yang-Mills field strength. Since string theory claims to
 give a consistent description of quantum gravity, solutions of the
 string effective action will contribute to our understanding of
 quantum gravity.

Not many exact solutions to the string equations of motion are
known. One of the
reasons for this is that knowledge about
the explicit form of  the higher order $\alpha'$
corrections to the string effective action have become
available  only
fairly recently \cite{Be2}.
 The higher order terms in $\alpha'$ in the
 effective action are for instance crucial
in the construction of the
five-brane soliton \cite{St1}.

In general, it appears difficult to find exact solutions to
the string equations of motion.
Fortunately, if one considers spacetimes with a symmetry,
 there exist transformations which generate new solutions from
 old ones.
 Examples of solution generating transformations specific
 to the string effective action are
 the target space duality
 transformations \cite{BUS,Ve1} and the
symmetry transformations of \cite{Se2}.

So far, the target space duality transformations have only been
derived
in a sigma model formulation of string theory \cite{BUS}.
The explicit
form of the duality transformation is only known to lowest
order in $\alpha'$. In general one expects that the duality
transformations will be modified with an infinite number
of terms of  increasingly higher order in $\alpha'$ but no
information is available about these higher order corrections.
Recently, Klim\v c\'\i k and Tseytlin found an exact duality
between some pp-waves
and flat space with non-vanishing antisymmetric tensor and dilaton
fields \cite{KT}.
A discussion, from the sigma model point of view, of other examples
of situations where the duality transformations are exact can be
found in \cite{KG}, \cite{GE}.

 A discussion of
less explicit examples of situations where the leading-order
duality
transformations do not acquire $\alpha'$ corrections  for a special
choice of
 field redefinitions can also be found in \cite{KT}.

The purpose of this paper is to find  duality transformations  which
form a
symmetry of the theory in the zero slope limit, and remain a symmetry
of the
theory even  with account taken of  $\alpha'$ corrections. However,
$\alpha'$
corrections include terms involving spin connection. This forced us
to formulate duality transformations not in terms of metric, as is
usually done,
but in terms of spin connection with torsion. The resulting
transformations
have a rather simple structure, which allowed us to
investigate higher
order  $\alpha'$ corrections.

We will also investigate special field
configurations for which these duality transformations
do  {\sl not} receive higher order corrections, i.e. are {\sl exact}
transformations to all orders in $\alpha'$. To be specific,
our starting point will be the supersymmetric string wave (SSW)
solutions to ten-dimensional string theory \cite{Be1}.
These solutions are characterized
by an  arbitrary vector field $A_\mu$.  It has been shown in
\cite{Ho4} that,
for the special case of plane fronted waves where only
$A_u\ne 0$, these wave
solutions are dual to the field outside a straight
fundamental string \cite{Da1}.  Furthermore, it is known that
 the plane fronted waves are an {\sl exact} solution
of the string equations of motion \cite{Gu1,Am1,Ho1}.
We will show  that the fundamental string  (FS) solution is also
an exact solution. As a consequence, we find
that the duality transformation, applied to
a plane fronted wave, is an exact transformation.

The above result can be generalized in the following way. We
will derive a theorem stating that, given a field
configuration that solves the zero slope limit of
the string equations, the same field configuration  can be
promoted to an
exact solution  provided
that  all $\alpha'$ corrections to the space-time supersymmetry
transformation rules vanish. The latter condition is
equivalent to the requirement that  for our solution (i)
 the (Lorentz + Yang-Mills)
Chern-Simons forms vanish and (ii) the so-called $T$-tensors (see
eqs.~(\ref{eq:t1})--(\ref{eq:t3})) vanish. In order to fulfill these
two conditions, it will be necessary in certain cases to
make a nontrivial Ansatz for the Yang-Mills gauge
fields\footnote{Note that the Yang-Mills vector fields do
not occur in the zero slope limit of the string effective action.}.

A corollary of our theorem is that, given two
solutions to the zero slope limit of the string equations,
which are connected to each other by a
duality transformations, then the duality transformation
is exact to all orders in $\alpha'$ provided both solutions
satisfy the two conditions in the theorem given above.
The main application of our theorem in this paper will be
to show that
the SSW solutions, after duality, lead to a {\sl generalized} FS
solution, which is again an exact solution.
Therefore, the duality transformation connecting the
two solutions is exact to all orders in $\alpha'$.

Our  approach is different from the one developed in \cite{Se2}  in
the
treatment of the duality transformation of vector fields. We do not
include
vector fields in the zero slope limit of the effective action but
treat the
non-abelian vector fields at the level of
$\alpha'$ corrections. These corrections cancel against
gravitational
$\alpha'$ corrections for the special configurations which we are
considering.

The organization of this paper is as follows. In section 2 we
will review the sigma model derivation of the target space
duality transformations in the zero slope limit. As a new result
we will also present the duality transformation of the
Yang-Mills gauge fields since we will need them in the following.
In section  3 we will explicitly show how the target
space duality invariance works in the zero slope limit. In the
next section this duality transformation will be
applied to construct the generalized FS solution. In section
5 we will give the derivation of our theorem.
In section 6 we will show that
the FS solution is an exact solution and hence that the
duality rotation connecting the SSW and generalized FS solution
is exact. Details about our notations and conventions
can be found in the Appendix A. Finally, in Appendices B and C we
prove that the generalized FS solution and the solution of \cite{KT}
are space-time supersymmetric.

\section {Sigma model duality}
We consider the sigma model action of the form
\begin{equation}
S = {1 \over 2\pi} \int d^2 z \;\left [(g_{\mu\nu} +
B_{\mu\nu})\;\partial
x^{\mu} \bar \partial x^{\nu} + i\;  \psi_I(\bar \partial
\psi^I +
 V_{\mu}{}^{IK} \bar \partial x^{\mu} \psi_K)\right ] \ .
\end{equation}
This action is a truncation of a supersymmetric sigma model \cite{HW}
related to the heterotic string. We assume that the background fields
$g_{\mu\nu},\; B_{\mu\nu}$ and $V_{\mu}{}^{IK}$ are
independent of
some coordinate $x$ and may depend on all remaining bosonic
coordinates
$x^\alpha$, where $\{x^\mu\} = \{x, x^\alpha \}$. This sigma model
allows
a straightforward generalization of the discrete target space duality
transformations rules \cite{BUS}
in presence of the vector fields in the background. We proceed in the
standard way \cite{RV} by presenting the first order action
\begin{eqnarray}
S_1& =& {1 \over 2\pi} \int d^2 z \Bigl[ g_{xx} A\bar A +
(g_{x\alpha}
+B_{x\alpha}) A\; \bar \partial x^{\alpha} + \Bigl((g_{\alpha x}
+B_{\alpha x })  \partial x^{\alpha} + V_x \Bigr)\bar A\nonumber\\
\nonumber\\
&+& (g_{\alpha\beta} + B_{\alpha\beta})\;\partial
x^{\alpha} \bar \partial x^{\beta} +
i\psi_I \bar \partial \psi^I +
V_{\alpha}\; \bar \partial
x^{\alpha}
+ \tilde \theta ( \partial \bar A - \bar \partial A) \Bigr] \ .
\end{eqnarray}
The following simplifying notation have been used
\begin{eqnarray}
V_x &\equiv &i \psi_I V_{x}{}^{IJ}  \psi_J\ , \nonumber\\
\nonumber\\
V_\alpha &\equiv &i \psi_I V_{\alpha}{}^{IJ} \psi_J\ .
\end{eqnarray}
By integrating out the Lagrange multiplier field $\tilde \theta$ on a
topologically trivial world-sheet, one recovers the original action
since the solution to the equation
$( \partial \bar A - \bar \partial A)=0$
is $A= \partial x, \bar A= \bar \partial x$. It is important to
stress that
this dualization procedure requires the non-vanishing value of
$g_{xx}$.
If one integrates out the gauge fields $A, \bar A$ one gets the
dual   model
\begin{equation}
\tilde S = {1 \over 2\pi} \int d^2 z \;\left [(\tilde g_{\mu\nu} +
\tilde B_{\mu\nu})\;\partial \tilde x^{\mu} \bar \partial \tilde
x^{\nu} + i\;
\psi_I(\bar \partial \psi^I +
 \tilde V_{\mu}{}^{IK} \bar \partial \tilde x^{\mu} \psi_K)\right ]
\ .
\end{equation}
The new coordinates are $\{\tilde x^\mu\} = \{\tilde \theta ,
x^\alpha \}$
and the dual metric and antisymmetric tensor field are
\begin{eqnarray}
\tilde g_{xx} & =& 1/g_{xx}\ , \qquad  \tilde g_{x\alpha} =
B_{x\alpha}/
g_{xx}\ , \nonumber\\
  \tilde g_{\alpha\beta} & =& g_{\alpha\beta} -
(g_{x\alpha}g_{x\beta} -
B_{x\alpha}B_{x\beta})/g_{xx}\ , \nonumber\\
  \tilde B_{x\alpha} & =& g_{x\alpha}/g_{xx}\ , \qquad
		\tilde  B_{\alpha\beta} = B_{\alpha\beta} +2
g_{x[\alpha}
B_{\beta]x}/g_{xx}\ .
\label{bus}\end{eqnarray}
The dual vector field is
\begin{equation}
\label{eq:dilaton}
 \nonumber\\
\tilde V_{x}{}^{JK}  = - V_{x}{}^{JK}  /g_{xx} \ , \qquad
\tilde V_{\alpha}{}^{JK} = V_{\alpha}{}^{JK} -
(g_{x\alpha} + B_{x\alpha} ) V_{x}{}^{JK} /g_{xx}\ .
\label{vec}\end{equation}
Taking into account the one-loop jacobian from integrating out $A,
\bar
A$-fields one finds, as usual, the dilaton transformation rules
\cite{BUS}\footnote{Strictly speaking, we should take
in (\ref{eq:dil}) the
absolute value $|g_{xx}|$ instead of $g_{xx}$.}
\begin{equation}
 \tilde \phi  = \phi - {1\over 2} \log g_{xx} \ .
\label{eq:dil}
\end{equation}
Thus by using the truncated version of a supersymmetric sigma model
it is easy to
supplement the well known target space duality transformations of the
metric and of the antisymmetric tensor field and of the dilaton by
the
accompanying transformations of the vector fields.

\section{Target-Space Duality in the Zero Slope Limit}

Our starting point is the bosonic part of the action of $N=1,\, d=10$
supergravity \cite{Ch1}\footnote{We use the same conventions as in
\cite{Be1},
except that we have redefined the axion field with $B_{\mu\nu}
\rightarrow -{3\over 2} B_{\mu\nu}$ in order to agree with the
duality transformations given in the previous section.
Further details about our notation and conventions can
be found in the appendix of \cite{Be1}.}:
\begin{equation}
\label{eq:action}
S(g_{\mu\nu}, B_{\mu\nu}, \phi)
={1\over 2}\int d^{10}x\ \sqrt {-g}e^{-2\phi}\biggl (
-R+4(\partial\phi)^2-
{3\over 4}H^2\biggr )\ ,
\end{equation}
where $g_{\mu\nu}$ is the metric, $\phi$ the dilaton and $H$  the
field strength of the axion $B_{\mu\nu}$.  We now consider the
special class of
field configurations which have an isometry generated by a Killing
vector
$k^\mu$. It is convenient to use a special coordinate system where
$k^\mu$
is constant and is  only nonzero in   one direction, let
us say the $x$ direction. Furthermore we assume that $k^\mu$ is a
non-null Killing  vector,
i.e.~$k^2\ne 0$. The isometry property then amounts to  the
following condition on the field configuration
$\{g_{\mu\nu}, B_{\mu\nu}, \phi\}$:
\begin{equation}
\label{eq:cond}
k^\mu\partial_\mu \{g_{\mu\nu}, B_{\mu\nu}, \phi\}
= \partial_x \{g_{\mu\nu}, B_{\mu\nu}, \phi\} = 0\ .
\end{equation}
We want to show that the action (\ref{eq:action}) is invariant
under
target space duality transformations modulo terms which contain the
derivative of one of the fields with respect to $x$. In other words
we
want to show that
\begin{equation}
S(g_{\mu\nu}, B_{\mu\nu}, \phi)=
S({\tilde g}_{\mu\nu}, {\tilde B}_{\mu\nu}, {\tilde \phi}) +
\int d^{10}x\ A(g,B,\phi) \partial_x B(g,B,\phi)\ ,
\end{equation}
where $A$ and $B$ are some  expressions in terms of
$g_{\mu\nu}, B_{\mu\nu}$ and $\phi$. If this is the case for {\sl
any}
field configuration that satisfies (\ref{eq:cond}) then the
target-space
duality transformation serves as a truly solution generating
transformation: the dual of {\sl any} solution of the field equations
that
is independent of the coordinate $x$ will automatically be another,
inequivalent,
solution of the same field equations.

In order to show that the action (\ref{eq:action}) is invariant under
the
duality transformations
given in the previous section it is convenient to use the zehnbein
instead of
the
metric.  The zehnbein transforms under the following
duality transformations:
\begin{eqnarray}
{\tilde e} _x^a &=& {1\over g_{xx}}e_x^a\ ,\nonumber\\
{\tilde e}_{\alpha}^a &=& e_\alpha^a - {1\over g_{xx}}
\bigl (g_{x\alpha} - B_{x\alpha}\bigr )e_x^a\ .
\label{zehnbein}\end{eqnarray}

{}From the point of view of reproducing the duality transformation of
the
metric, given in equation (\ref{bus}), we could equally well use a
different
duality transformation for the zehnbeins. We could have used the
transformation
\begin{equation}
{\tilde e} _x^a = - {1\over g_{xx}}e_x^a\ ,\qquad
{\tilde e}_{\alpha}^a = e_\alpha^a - {1\over g_{xx}}
\bigl (g_{x\alpha} + B_{x\alpha}\bigr )e_x^a\ .
\end{equation}
This transformation coincides exactly with the duality transformation
of the
vector field
in equation (\ref{vec}). We have found, however,  that  to provide
the absence
of
$\alpha'$ corrections we have to use the duality transformation of
the zehnbein
given in equation (\ref{zehnbein}) if the one for the vector field is
given in
equation
(\ref{vec}).  Another possibility is to change both of them to the
opposite
one. This will be a necessary condition for the duality symmetry to
preserve
the embedding condition of the spin connection into a subgroup of the
gauge
group.

Under a duality transformation the
spin-connection and the axion field strength transform as:
\begin{eqnarray}
{\tilde \omega}_c{}^{ab} &=& \omega_c{}^{ab} - {1\over k^2}k_ck^d
\Omega_{d-}{}^{ab} -{2\over k^2}k^dk^{[a}\Omega_{d-,c}{}^{b]}\ ,
\nonumber\\
{\tilde H}_{abc} &=& H_{abc} -{2\over k^2}k^dk_{[a}
\Omega_{d-,bc]}\ ,
\end{eqnarray}
where  the torsionful spin connections $\Omega_{\mu \pm }{}^{ab}$ are
defined by\footnote{Note that with this definition the $\Omega_\pm$
of this paper is identical to the $\Omega_\pm$ of \cite{Be1} {\sl
after}
the redefinition $B_{\mu\nu} \rightarrow - {3\over 2}B_{\mu\nu}$.}
\begin{equation}
\label{eq:torsion}
\Omega_{\mu\pm}{} ^{ab} = \omega_\mu{}^{ab}(e) \mp {3\over 2}
H_\mu{}^{ab}\ .
\end{equation}
For later convenience we note that  the dual of the torsionful spin
connections are given by
\begin{eqnarray}
\label{eq:dualtorsion}
{\tilde \Omega}_{c-}{}^{ab} &=& \Omega_{c-}{}^{ab} - {2\over
k^2}k_ck^d
\Omega_{d-}{}^{ab}\ ,\nonumber\\
{\tilde \Omega}_{c+}{}^{ab} &=& \Omega_{c+}{}^{ab} - {4\over k^2}k^d
k^{[a}
\Omega_{d-,c}{}^{b]}\ .
\label{spin}\end{eqnarray}
Using the identity
\begin{equation}\label{eq:Palatini}
- \int d^{10}x\ \sqrt {-g}e^{-2\phi}
R \, =\, \int d^{10}x\ \sqrt {-g}e^{-2\phi}\biggl (
\omega_a{}^{ac}\omega_b{}^{bc} + \omega_a{}^{bc}\omega_b{}^{ca}
+ 4 (\partial_a\phi)\omega_b{}^{ba}
 \biggr ) \ ,
\end{equation}
we can now rewrite the action (\ref{eq:action})
in the following convenient form:
\begin{equation}
\label{eq:action2}
S={1\over 2}\int d^{10}x\ ee^{-2\phi}\biggl (
4(\partial^a\phi + {1\over 2}\omega_b{}^{ba})^2 +
\omega_a{}^{bc}\omega_b{}^{ca} - {3\over 4} H_{abc}H^{abc}
\biggr )\ ,
\end{equation}
where $e= {\rm det}\ e_\mu^a$.
Using the additional identity
\begin{equation}
k^bk^c\omega_{b,ca} = -{1\over 2}\partial_a k^2\ ,
\end{equation}
and  transformation rule
\begin{equation}
\tilde e = {1\over k^2} e\ ,
\end{equation}
it is straightforward to show that the action  (\ref{eq:action2})
is invariant under target-space duality transformations.  We note
that
the action consists of three parts which are separately duality
invariant:
the combinations  $ee^{-2\phi}, \ (\partial^a\phi +
 {1\over 2}\omega_b{}^{ba})$ and $(\omega_a{}^{bc}\omega_b{}^{ca} -
 {3\over 4} H_{abc}H^{abc})$ are all  three  duality invariant.

Thus we have shown that the target space action in the zero slope
limit is
invariant under the sigma model duality transformations, given in
equations
(\ref{bus}) and (\ref{eq:dil}). The vector fields are not present in
the
effective action in the zero slope limit.

\section{New Solutions in the Zero Slope Limit}

The purpose of this section is to show how the duality
transformations
can be used to generate new solutions in the zero slope limit.
Higher-order $\alpha'$ corrections will be considered in the next two
sections.
Our starting point is the Supersymmetric
String Wave (SSW) solution of \cite{Be1}. In \cite{Be1}
it was shown that the
SSW solves the string equations of motion to all orders in $\alpha'$.
Here we will only consider the zero slope limit. In particular, we
will set
the vector gauge fields equal to zero. The solution thereby reduces
to the
one given in \cite{Ru1}, \cite{Ho3}.
A crucial  feature of the SSW
solution is the existence of a null Killing vector $l^\mu$ with
$l^2=0$.
This Killing vector generates an isometry in the $v$ direction where
we use light-cone coordinates $x^\mu = (u,v,x^i)\ (i=1,\dots ,8)$.
Since this Killing vector is null we cannot use it for a
duality transformation. We therefore make the additional assumption
that the
fields occurring in the SSW solution are also independent of the $u$
coordinate. Since the Ansatz for the dilaton in the SSW solution only
depends on $u$,  it must be a constant and for simplicity we
will take this constant to be zero.  The remaining nonzero fields
$g_{\mu\nu}$ and $B_{\mu\nu}$
are both described in terms of one vector function
of the transverse coordinates $x^i$:
\begin{equation}
A_\mu(x^i) = \{A_u(x^i), A_v=0, A_i(x^i)\}\ .
\end{equation}
They are given by the Brinkmann metric \cite{Br1}
and the following 2-form\footnote{We use the following form notation:
$ds^2 = g_{\mu\nu}dx^\mu dx^\nu$ and $B = B_{\mu\nu}dx^\mu\wedge
dx^\nu$.}:
\begin{eqnarray}
\label{eq:SSW}
ds^2 &=& 2dudv + 2A_\mu dx^\mu du - dx^idx^i\ ,\nonumber \\
\label{eq:SSW2}
B &=& 2 A_\mu dx^\mu \wedge du\ .
\end{eqnarray}
The equations that $A_u(x^i)$ and $A_i(x^i)$ have to satisfy are:
\begin{equation}
\label{eq:Lapl}
\triangle A_u = 0\ , \hskip 1.5truecm \triangle\partial^{[i}A^{j]} =
0\ ,
\end{equation}
where the Laplacian is taken over the transverse directions only.
Since $g_{\mu\nu}$ and $B_{\mu\nu}$ are independent of
$u$ and $v$ they are  independent of $x$ and $t$.
For our duality transformation we
will use the isometry in the space-like $x$ direction. The non-null
Killing vector generating this isometry will be denoted by $k^\mu$.
Note that we have now two isometries given by
\begin{equation}
l^\mu\partial_\mu \{g_{\mu\nu}, B_{\mu\nu}\} = \partial_v
\{g_{\mu\nu}, B_{\mu\nu}\} = 0 \ ,\hskip 1truecm
k^\mu\partial_\mu \{g_{\mu\nu}, B_{\mu\nu}\} = \partial_x
\{g_{\mu\nu}, B_{\mu\nu}\} = 0\ .
\end{equation}
A straightforward application of the sigma-model duality
transformations
given in (\ref{bus}), (\ref{eq:dil}) on the  SSW solution given in
eq.~(\ref{eq:SSW})
leads to the following new solution of the
zero slope limit equations of motion:
\begin{eqnarray}
\label{eq:new}
ds^2 &=&  - (A_u-1)^{-1}\bigl \{ 2dudv + 2 A_i dudx^i
\bigr \} - dx^idx^i\ ,\nonumber\\
B &=& (A_u-1)^{-1} \bigl
\{ 2A_u du \wedge dv + 2 A_idu \wedge dx^i \bigr \}\ ,\\
\phi  &=& -{1\over 2} {\rm log} (A_u -1)\ .\nonumber
\end{eqnarray}
Note that we can make the following particular choice of the vector
function $A_\mu$\footnote{In order to solve the equations
(\ref{eq:Lapl}),
it is understood that a source term at $r=0$, representing a
fundamental string, has been added to these equations.}:
\begin{equation}
A_u = -{\tilde c M\over r^6}\ , \hskip 1.5truecm A_i = 0\ ,
\end{equation}
where $r^2 = x^ix_i$ and $\tilde c$ a constant.  The solution given
in
(\ref{eq:new}) reduces then to the solution of
\cite{Da1} corresponding to the field outside a fundamental string
with total mass $M$. We will hence refer to the solution
(\ref{eq:new})
 as the generalized FS solution\footnote{Note that the solution
(\ref{eq:new})
does not yet include the vector fields. The complete generalized
FS solution, including the vector fields, is given in
(\ref{eq:generalised}).}.

It was shown in \cite{Be1} that the SSW solution is supersymmetric
under 8 of the 16 ten-dimensional supersymmetries.
In  \cite{Da1}, it has been shown that  the dual FS solution,
for the special case that $A_i=0$, is  again supersymmetric. In
Appendix B
 we will show that
also the generalized FS solution, with $A_i\ne 0$,  is
supersymmetric.

\section{ $\alpha'$ corrections}

In this section we will consider the $\alpha'$ corrections to the
zero
slope limit. In particular, we will derive a theorem stating that
any solution to the zero slope limit string equations
can be promoted to an exact solution to all orders in $\alpha'$
provided that (i) the $T$-tensors defined by eqs.~(\ref{eq:t1}) -
(\ref{eq:t3}) vanish and (ii) a particular combination of
the (Lorentz + Yang-Mills) Chern-Simons terms vanish.
We note that the discussion in this section is independent of any
particular
solution to the field equations.

There exists one remarkable property of the duality
transformations considered above,  which is of crucial importance for
the
understanding of $\alpha'$ corrections. This property is best see
in terms of the duality transformations of  the torsionful spin
connections
$\Omega_{c-}{}^{ab}$, defined above in eq.~(\ref{eq:torsion}).  The
useful
form of this
transformation is
\begin{equation}
\label{eq:tro}
{\tilde \Omega}_{c-}{}^{ab} = \Pi _c{}^d \; \Omega_{d-}{}^{ab} \ ,
\end{equation}
where we have introduced the projector
\begin{equation}
\Pi _c{}^d \equiv \delta _c{}^d -  {2\over k^2}k_ck^d \ .
\end{equation}
The square of this projector  is a unit operator:
\begin{equation}
\Pi _c{}^d \Pi _d{}^e = \delta _c{}^e \ .
\end{equation}
This property of the projector nicely confirms the fact that we are
performing
a discrete operation and that two such operations bring us
back to the
original
configuration
since
\begin{equation}
{\tilde {\tilde \Omega}}_{c-}{}^{ab} = \Pi _c{}^d \;  {\tilde
\Omega}_{d-}{}^{ab} =
\Pi _c{}^d \; \Pi _d{}^e \; \Omega_{e-}{}^{ab} =
\Omega_{c-}{}^{ab} \ .
\end{equation}

 Now let us consider the $\alpha^{\prime}$ corrections to  the string
effective action.  It is well known that one has to add
to $S^{(0)}$ the Lorentz and Yang-Mills Chern-Simons terms which play
a
crucial role in the Green-Schwarz anomaly cancellation
mechanism\footnote{
Not much is known about the properties of $\alpha^{\prime}$
corrections, which are not related to anomalies.
For some recent results about these additional corrections, see
\cite{deRoo} and references therein. We will not consider these
corrections in this paper.}.  These terms
break supersymmetry. To restore supersymmetry order by order in
$\alpha^{\prime}$, one has to add to $S^{(0)}$  an infinite series
of higher
order
terms in $\alpha^{\prime}$.
  By the
procedure of adding terms to restore supersymmetry, the
effective action was obtained in \cite{Be1}
up to order $O(\alpha^{\prime 4})$ terms:
 \begin{eqnarray}\label{eric}
S & = & {1\over 2}\int d^{10}x\ \sqrt {-g}e^{-2\phi}\biggl (
-R+4(\partial\phi)^2- {3\over 4}H^2+ \nonumber \\ &&+{1\over 2}T +
2\,
\alpha^{\prime} T^{\mu\nu} T_{\mu\nu}
 +6 \, \alpha^{\prime} T^{\mu\nu\lambda\rho}
T_{\mu\nu\lambda\rho}+ O(\alpha^{\prime 4})\biggr )\ ,
\end{eqnarray}
where the antisymmetric tensor
$T_{\mu\nu\lambda\rho}$, the symmetric tensor $T_{\mu\nu}$, and
the scalar $T$ are given by
\begin{eqnarray}
T_{\mu\nu\lambda\rho} &=&2  \alpha^{\prime} \biggl ( \,
R_{[\mu\nu}{}^{ab}\bigl
(\Omega_-)
                          \, R_{\lambda\rho]}{}^{ab}\bigl (\Omega_-)
+
                           \frac{1}{30}\, {\rm tr} F_{[\mu\nu}
                           F_{\lambda\rho]} \biggr )\label{eq:t1} \
,\\
T_{\mu\nu} &= &2  \alpha^{\prime} \, \biggl (
R_{\mu\lambda}{}^{ab}\bigl
                         (\Omega_-)
                R^{\lambda } {}_{ \nu} {}^{  ab}(\Omega_-)
                + \frac{1}{30} \, {\rm tr} F_{\mu  \lambda}
                F^{\lambda}{}_{\nu} \biggr )\label{eq:t2} \ ,\\
T & = & T_\mu{}^\mu \ \ .
   \label{eq:t3}
\end{eqnarray}
In the above expression there are explicit {\it and} implicit
$\alpha^{\prime}$
corrections. The explicit corrections always appear via $T$-tensors,
and  they
are
essentially the $\alpha^{\prime}$ factors in front of eqs.
(\ref{eq:t1})-(\ref{eq:t3}).
The implicit $\alpha^{\prime}$ corrections always appear via the
torsion $H$
which is
defined by the following iterative procedure: At
the lowest order $H$ is just $H^{(0)}_{[\mu\nu\rho]}=
\partial_{[\mu}B_{\nu\rho]}$. With $H^{(0)}$ we calculate the lowest
order
$\Omega_{\pm}=\Omega_{\pm}^{(0)}$, as given in
eq.~(\ref{eq:torsion}).
  At
first order in $\alpha^{\prime}$, $H=H^{(1)}$ is $H^{(0)}$ corrected
by the
Yang-Mills Chern-Simons term  and the
Lorentz Chern-Simons term corresponding to the zero-order
$\Omega_{-}=\Omega_{-}^{(0)}$:
\begin{equation}
H^{(1)}_{\mu\nu\rho}  = \partial_{[\mu}B_{\nu\rho]}
+\alpha^{\prime}( \omega_{\mu\nu\rho}^{Y.M.} +
\omega_{\mu\nu\rho}^{L}) \ ,
\end{equation}
where
\begin{equation}
\omega_{\mu\nu\rho}^{Y.M.}= -\frac{1}{5}
{\rm tr}
\biggl \{ V_{[\mu}\partial_\nu V_{\rho]}-
{1\over 3}V_{[\mu}V_\nu V_{\rho]}\biggr \} \ ,
\end{equation}
and
\begin{equation}
\label{eq:LCS}
\omega_{\mu\nu\rho}^{L} = -6 \,  \biggl
 \{ \Omega^{(0)}_{[\mu-}{}^{ab}\partial_\nu
\Omega^{(0)}_{\rho]-}{}^{ab}-
{1\over 3}\Omega^{(0)}_{[\mu-}{}^{ab}\Omega^{(0)}_{\nu-}{}^{ac}
\Omega^{(0)}_{\rho-]}{}^{cb}\biggr \} \ .
\end{equation}

With $H^{(1)}$ one would get $\Omega^{(1)}$ using again its
definition eq.
(\ref{eq:torsion}) and $H^{(2)}$ would be given by the above
expression but
with
$\Omega^{(0)}$ replaced by $\Omega^{(1)}$. Iterating this procedure
one gets
the
all-order expression $H$ for the torsion which involves the promised
infinite
series of
corrections.

In short, to be able to understand the properties of
$\alpha^{\prime}$
corrections to specific configurations we have to  calculate the
value of the Lorentz and Yang-Mills Chern-Simons term and the values
of
all
$T$-tensors, presented above.

To study the  corrections to the equations of motion we will use the
corresponding analysis, performed in \cite{Be1}.
 We
first have to vary the action (\ref{eric}) over all the fields
present in the
theory,
and  only then substitute the solutions in the corrected equations.
We will
study the linear corrections separately. The equations of motion
corrected
up to first order come from the action
\begin{equation}
S^{(1)}={1\over 2}\int d^{10}x\ \sqrt
{-g}e^{-2\phi}\biggl (-R+4(\partial\phi)^{2}
-{3\over 4}H^{2} +\frac{1}{2}T \biggr )\ .
 \end{equation}
 All terms of order
 $\alpha^{\prime 2}$ and
higher are
neglected at this stage.
The corrections linear in $\alpha^{\prime}$ to the lowest order
equations of
motion are
derived from the variation  $\delta (S^{(1)}-S^{(0)})\equiv \delta
\Delta S$.
It is
convenient to perform this variation with respect to the explicit
dependence
of
the action on
$g_{\mu\nu}$, $V_{\mu}$, $\phi$ and $B_{\mu\nu}$, and then with
respect
to
the implicit dependence on these fields through the torsionful spin
connection
$\Omega_{-}$, that is
 \begin{eqnarray}
\delta \Delta S & = &
\frac{\delta\Delta S}{\delta g_{\mu\nu}} \delta g_{\mu\nu}
+\frac{\delta\Delta S}{\delta B_{\mu\nu}}\delta B_{\mu\nu}
+\frac{\delta\Delta S}{\delta\phi}\delta\phi+ \nonumber \\
&&+\frac{\delta\Delta S}{\delta V_{\mu}}\delta V_{\mu}
+\frac{\delta\Delta S}{\delta \Omega_{\mu -}{}^{ab}}
\delta\Omega_{\mu -}{}^{ab}\ ,\label{DS}
\end{eqnarray}
where
\begin{equation}
\delta\Omega_{\mu -}{}^{ab}=
\frac{\delta\Omega_{\mu -}{}^{ab}}{\delta g_{\mu\nu}} \delta
g_{\mu\nu}
+\frac{\delta\Omega_{\mu -}{}^{ab}}{\delta B_{\mu\nu}}\delta
B_{\mu\nu}
+\frac{\delta\Omega_{\mu -}{}^{ab}}{\delta V_{\mu}}\delta V_{\mu}\ .
\end{equation}
The explicit variations are
\begin{eqnarray}
\frac{\delta\Delta S}{\delta g_{\mu\nu}} & = &
-\frac{1}{4}\sqrt{-g}e^{-2\phi}(T^{\mu\nu}-g^{\mu\nu}T)\ ,
\\
\frac{\delta\Delta S}{\delta \phi} & = &
-\frac{1}{2}\sqrt{-g}e^{-2\phi}T\ ,  \\
\frac{\delta\Delta S}{\delta B_{\mu\nu}}
& = & \frac{3}{4}\,
\partial_{\lambda}[\sqrt{-g}e^{-2\phi}
(H^{(1)}{}^{\lambda\mu\nu}
-H^{(0)}{}^{\lambda\mu\nu})] \ ,     \\
\frac{\delta\Delta S}{\delta V_{\mu}} & =
&\frac{1}{15}\alpha^{\prime}\biggl\{
                                        \partial_\lambda
                                     \bigl  ( \sqrt{-g}e^{-2\phi}
F^{\lambda\mu}\bigr
                                      )
+\sqrt{-g}e^{-2\phi}[V_\lambda
,F^{\lambda \mu}]+
                                         \nonumber \\
                                 &&-\frac{3}{2}  \sqrt{-g}e^{-2\phi}
H^{(0)\mu}{}_{\lambda\rho}
                                 F^{\lambda\rho}
+\frac{3}{2}V_{\rho}
                               \partial_{\lambda}(\sqrt{-g}e^{-2\phi}
                                 H^{(0)}{}^{\lambda\mu\rho})\biggr
\}\ .
\end{eqnarray}
If some solution of the zero slope limit equations has the property
that the
$T$-tensors vanish on this solution, the first two equations above do
not
get linear $\alpha'$ corrections. The third equation is not
corrected,
provided the combination of Lorentz and Yang-Mills Chern-Simons term
is vanishing for the configuration. The last equation is satisfied if
the
classical equation of motion for the vector field\footnote{One can
include the
vector field action in the classical Lagrangian of supergravity
interacting
with the Yang-Mills multiplet. We treated the vector field here as
coming at
the level of $\alpha'$ corrections, which is natural in the framework
of
string effective action.}
 as well as the equation of motion for the $B_{\mu\nu}$-field in the
zero
slope limit are satisfied.
We next consider the implicit variations represented by the last term
in eq.
(\ref{DS}). In \cite{Be2} a rather non-trivial
property of
the  string effective action was analysed. This property is  that
all
the
implicit variations are proportional to the lowest order equations of
motion
of
the
different fields\footnote{For more details, see
\cite{Be2} and the Appendix of \cite{Be1}.}.
 We therefore conclude that, to linear order in $\alpha'$,
any solution of the
lowest
order equations of motion are also solutions of the equations of
motion
corrected to
order $\alpha^{\prime}$ provided that (i) the T-tensor $T_{\mu\nu}$
and (ii) the (Lorentz + Yang-Mills) Chern-Simons terms vanish.

 Now let us consider the higher order $\alpha'$ corrections. The
general
structure of the
bosonic part of the effective action, which can be obtained by the
procedure
outlined before based upon the restoration of supersymmetry, has been
conjectured in ref. \cite{Be1}. New terms in the action are quadratic
or of
 higher degree  in the $T$-tensors. Therefore their contribution to
the string equations
of motion automatically vanishes for the configurations with
vanishing
$T$-tensors. This concludes the proof of the theorem stated at the
beginning of this section.

Finally, we would like to note that the vanishing of a
combination of Lorentz and Yang-Mills Chern-Simons term in
$H_{\mu \nu \lambda}$ and the vanishing of  all $T$-tensors
is sufficient for the absence of
corrections to the supersymmetry transformation laws and the Killing
spinors \cite{Be1}, \cite{Be2}.
As a corollary of this we conclude that
the condition for a field configuration that solves
the zero slope limit equations of motion to be exact, i.e. to
be free of
$\alpha^{\prime}$-corrections, coincides with the property of the
configuration to have vanishing $\alpha^{\prime}$-corrections to
classical supersymmetry transformations.

\section{Exact duality between pp-waves and strings}

The complete discussion of the supersymmetric string waves and
$\alpha'$ corrections has been performed in \cite{Be1}.
Our final conclusion  was that the on-shell action, the  fields that
solve
the lowest order equations of motion and the Killing spinors for the
SSW
solutions do not
receive any higher order string corrections.

In this section we would like to use the pp-wave solution,
i.e.~the SSW solution with $A_i=0$, as the starting point for
investigation of the corrections to duality symmetry in the target
space. In the next section we will consider the general case with
$A_i \ne 0$.

Gravitational plane fronted waves \cite{Br1} have a null Killing
vector
$\nabla_{\mu} l_{\nu} = 0\ ,  \quad l^{\nu}l_{\nu}= 0$ and
 very special dependence of this null vector. This
simplifies the analysis of higher order corrections to field
equations.

Consider the class of 10-dimensional pp-waves with
metrics
of the form  \begin{equation}\label{d}
ds^2 = 2 du dv + K (u, x^i ) du^2 - dx^i d x^i \ ,
\end{equation}
where $i = 1,2, ..., 8$, the Riemann curvature is \cite{Ho1,Ho3}
\begin{equation}
R_{\mu\nu\rho\sigma} = - 2 l_{[\mu}( \partial_{\nu]}
\partial_{[\rho} K )
l_{\sigma]}\ .
\end{equation}
The curvature is orthogonal to $l_{\mu}$ in all its indices.
This fact is of crucial importance in establishing that all higher
order in
$\alpha^{\prime}$ terms in the equations of motion are zero due to
the
vanishing of all the
possible contractions of curvature tensors.
The dilaton, the antisymmetric tensor field and the
vector field are absent in this solution. The only non-trivial
function in the
metric has to satisfy the equation
\begin{equation}
\triangle K = 0\ ,
\end{equation}
where
 $\triangle=\partial_{i}\partial_{i}$ is the flat space
Laplacian.
The spin connection of these pp-waves is given by the following
expression:
\begin{equation}
\omega_\mu{}^{ab}  =  -  l_\mu \; l^{[a}\partial^{b]} K \ .
\end{equation}
Note also that the
indices $ab$ related to the fact that the spin connection is a
Lorentz-Lie-algebra valued object\begin{equation}
\omega \equiv dx^\mu \omega_\mu {} ^{ab} M_{ab}=
- dx^\mu l_\mu \;l^{[a}\partial^{b]} K \; M_{ab}
\end{equation}
have at least one null vector.
In this case the spin connection coincides with the torsionful spin
connection
since the antisymmetric field strength tensor $H_{abc}$  vanishes:
\begin{equation}
 \Omega_{c-}{}^{ab}= -  l_c \; l^{[a}\partial^{b]} K\ .
\end{equation}
This spin connection is orthogonal to $l_a$ in all indices and what
is very
important, has at least one $l$-vector in $ab$-type indices.

In the discussion below we also have to take into
account that
\begin{equation}
\partial_\mu l^a =0\ .
\end{equation}
The structures
which we need, Lorentz Chern-Simons term  given in
eq.~(\ref{eq:LCS}) and all T-tensors,
given in eqs.~(\ref{eq:t1})-(\ref{eq:t3}),
always have all
$ab$-type
indices contracted. This means that for the pp-waves all these
structures
vanish.
There is no way to contract those null vectors, if they are
contracted with
another $l$ we get zero and if they are contracted with $\partial$ we
have
zero again, since this a Killing condition, the solution is
independent of $v$.
Thus for pp-waves the $\alpha^{\prime }$
corrections
to the action, to the equation of motion and to supersymmetric
transformations of all fermion fields  vanish.

As explained in \cite{Ho4} the dual version of pp-waves (without a
dilaton, antisymmetric field and vector field) is a string
solution of \cite{Da1}. As explained above, the pp-waves have no
$\alpha'$ corrections since the Lorentz Chern-Simons term  and all
T-tensors
vanish for pp-waves. Now that we have a simple duality transformation
rule
for the torsionful spin connection we can investigate the
corresponding
features
of the solution dual to the pp-waves.

As stated in the beginning of the previous section,
the duality transformation
${\tilde \Omega}_{c-}{}^{ab} = \Pi _c{}^d \; \Omega_{d-}{}^{ab}$ does
not
affect the structure of the $ab$-type indices
in this expression. Therefore the proof that the Lorentz Chern-Simons
term
and all T-tensors vanish for pp-waves is immediately extended to the
dual
version, i.e.~to the string solution of \cite{Da1}. The reason for
that is
that all
those
terms always have at least one null Killing vector in $a$ or $b$
direction,
which has to be contracted with another $a$- or $b$-type index.
Although
the contraction with new Killing vector $k^\mu$ is possible in
principle and
would give a  non-vanishing contribution since $k^2\ne 0$ and
$k\cdot l \ne 0$, we see from the structure
of
Chern-Simons term  and  T-tensors
 that after duality transformation no such contraction actually
occurs.

Having established the vanishing of $\alpha'$ corrections to Lorentz
Chern-Simons term
and all T-tensors for the FS string solution of \cite{Da1}
we may apply the general
analysis performed in the previous section. We find that there are
no
corrections to the FS solution of the zero slope limit equations.
We therefore conclude that both the FS solution of \cite{Da1}
and the pp-waves are exact solutions of the string effective action.

A simplifying feature of the pp-waves is the fact that they cannot
have
corrections independently of the specific form of the $\alpha'$
corrections. For
the dual version, the FS solution, we have under complete control
only the
$\alpha'$ corrections described above which are related to anomalies.
{\sl A priori} they could receive an infinite set of corrections, in
principle,
unless the
iterative
procedure requires them to vanish on each step as we have seen above.

The conclusion of this section is that
the duality transformation, which relates the pp-waves to the FS
solution
is an exact duality transformation, i.e.~the $\alpha'$ corrections
to the duality transformation vanish for this case.

\section{Exact duality between supersymmetric string waves and
generalized FS solutions}

In this section we  extend the discussion of the previous section
to the generalized FS solution given in
eq.~(\ref{eq:new}) for the case that $A_i \ne 0$.  It has been shown
in \cite{Be1} that even for this case the corresponding SSW solution
is an
exact solution to all orders in $\alpha'$. This is a nontrivial
result since with $A_i\ne 0$ both the Lorentz Chern-Simons term
as well as the $T$ tensors do {\sl not} vanish. To make both vanish
one must
make a nontrivial Ansatz for the vector gauge fields $V_\mu{}^{IJ}$
and
embed the torsionful spin connection $\Omega_-$ in an $SO(8)$
subgroup
of the gauge group, thereby identifying the spin connection
with the gauge field:
\begin{equation}
\label{eq:embedding}
{1\over \sqrt{30}}V_\mu{}^{IJ}
=  l_\mu V^{IJ}\hskip .5truecm \equiv \hskip .5truecm
\Omega_{\mu-}{}^{ab} = l_\mu {\cal A}^{ab}
\ \ \ \ (a,b,I,J=1,\dots ,8)\  .
\end{equation}
Here the Yang-Mills indices refer to the adjoint representation of
$SO(8)$.   One then makes use of the fact that
with a non-zero gauge field the Lorentz Chern-Simons term
always occurs together with a Yang-Mills Chern-Simons term. The
same applies to the $T$ tensors where the $R^2$ terms are
always accompanied by similar $F^2$ terms.
The above identification then leads to a  cancellation
between the spin connection  and gauge field terms such that even
with $A_i\ne 0$ the (Lorentz + Yang-Mills ) Chern-Simons terms
and the (generalized) $T$ tensors do vanish. Of course this
cancellation
only involves terms with the  function $A_i$. The terms
involving
$A_u$  in the Lorentz sector already cancel by themselves as has
been shown in the previous section.

To investigate what happens with the  (Lorentz + Yang-Mills)
Chern-Simons
terms and the $T$ tensors for the generalized FS solution
we apply a duality transformation on the torsionful spin connection
corresponding to the SSW solution. This  leads to the following
expression of $\Omega_-$ for the generalized FS solution:
\begin{equation}
\tilde {\Omega}_{c-}{}^{ab} = \{l_c -
 {2 k\cdot l\over k^2}k_c\}{\cal A}^{ab} = \Pi_c{}^d l_d {\cal
A}^{ab}\
{}.
\end{equation}
Note  that the structure of the $ab$ indices
remains unchanged under a duality rotation.

In order to show that
the generalized FS solution is again a solution of the
field equations to all orders in $\alpha'$ it is essential that the
above-mentioned cancellation between the spin connection and gauge
field terms
in the (Lorentz + Yang-Mills) Chern-Simons terms and the $T$ tensors
is not spoiled by the duality transformation. We therefore require
that
the embedding is duality invariant.
In order to obtain a duality-invariant embedding
we want that $\Omega_-$  transforms in the same way as the Yang-Mills
gauge fields. Luckily enough it turns out that this is indeed the
case. Using the duality transformation of the vector fields given in
(\ref{eq:dilaton}) one can show that the duality transformations
of the spin connection and the gauge fields have precisely the same
form:\footnote{
In the equations given below it is understood that the SSW metric is
used.
Note that the dual gauge fields are given with flat indices. To
convert
them into curved indices one should use the metric corresponding to
the FS
solution.}:
\begin{equation}
\tilde {\Omega}_{c -}{}^{ab} = \Pi_c{}^d l_d {\cal A}^{ab}\hskip
1truecm
{\underline  {{\rm and}}} \hskip 1truecm
{1\over \sqrt{30}}\tilde {V}_c{}^{IJ} = \Pi_c{}^d l_d {\cal A}^{IJ}\
{}.
\end{equation}
This means that the embedding condition (\ref{eq:embedding}) is
indeed
duality invariant and the cancellations which took place between
the spin connection and gauge field terms in the SSW solution
again take place, {\sl after} the duality rotation,  for the
generalized
FS solution. Hence, for the generalized FS solution we
can again derive that the (Lorentz + Yang-Mills) Chern-Simons terms
and the $T$-tensors are zero.
At this point we can use the results of section 5 where
we have shown that the vanishing of the (Lorentz + Yang-Mills)
Chern-Simons
form and the $T$ tensors is enough to ensure that the zero-slope
limit
solution extends to a solution to all orders in $\alpha'$.

The conclusion is therefore that the generalized FS solution
given by
\begin{eqnarray}
\label{eq:generalised}
ds^2 &=&  - (A_u-1)^{-1}\bigl \{ 2dudv + 2 A_i dudx^i
\bigr \} - dx^idx^i\ ,\nonumber\\
B &=& (A_u-1)^{-1} \bigl
\{ 2A_u du \wedge dv + 2 A_idu \wedge dx^i \bigr \}\ ,\\
\phi  &=& -{1\over 2} {\rm log} (A_u -1)\ ,\nonumber\\
V_\mu^{IJ} &=&  - (A_u-1)^{-1} l_\mu {\cal A}^{IJ}\hskip 1truecm
(I,J=1,\dots ,8)\ ,\nonumber
 \end{eqnarray}
 with ${\cal A}_{\mu\nu}=\partial_\mu A_\nu -
\partial_\nu  A_\mu$ solves the string equations of motion to
all orders in $\alpha'$.
The equations that $A_u(x^i)$ and $A_i(x^i)$ have to satisfy are:
\begin{equation}
\triangle A_u = 0\ , \hskip 1.5truecm \triangle\partial^{[i}A^{j]} =
0\ ,
\end{equation}
where the Laplacian is taken over the transverse directions.
Furthermore, the duality transformation  connecting the SSW solution
and the generalized FS solution is exact to all orders in
$\alpha'$.

\section{Conclusion}

In this paper we   presented a set of duality transformations for
the
zehnbein,
spin connection (with torsion) and vector fields. They are given in
equations
(\ref{bus}), (\ref{vec}), (\ref{zehnbein}), (\ref{spin}). The most
elegant
duality transformation of the spin connection with torsion is given
in equation
 (\ref{eq:tro}).
This specific transformation plays a crucial role in the analysis of
$\alpha'$
corrections.

We have found the conditions under which a target space
duality
symmetry
is exact, i.e.~does not acquire $\alpha'$ corrections.
Two configurations can be qualified as being exactly dual to each
other if:

i) there exists a non-null Killing vector in the original as well as
in the dual
configuration,
which allows one to identify the corresponding sigma-model duality
transformation. This transformation defines a symmetry of the zero
slope
limit of the string effective action in the target space.

ii) the condition for the action and equations of motion not to
acquire
$\alpha'$ corrections is provided by the vanishing of the combination
of the
Lorentz and Yang-Mills Chern-Simons term as well as by the vanishing
of  the
$T$-tensors for the  original as well as for the dual
configuration.

iii) exact duality in all explicit examples known to us brings one
supersymmetric
configuration to another supersymmetric configuration. The zero slope
supersymmetric transformation rules are not affected by the
$\alpha'$ corrections for the original as well as the final
configurations
related by exact duality.

Exact duality  defined above serves as an exact solution generating
transformation. As an example of configurations related by exact
duality we
have analysed the pp-waves and fundamental string solutions
\cite{Da1}.
Both solutions are free of $\alpha'$ corrections \footnote {As
explained
above, in the string case we have control only
on anomaly-related
$\alpha'$ corrections.}.

As a further application of our general results we have also checked
that the examples of exact duality  investigated
by Klim\v c\' \i k and Tseytlin \cite{KT}
also have some unbroken space-time supersymmetry. Their original
configuration is supersymmetric, since it is equivalent to one of the
G\"uven \cite{Gu1}
solutions. The  dual configuration turns out to be also
supersymmetric, as
we have shown in Appendix C.

A more general example of exact duality is given by
the supersymmetric string waves \cite{Be1} and the generalized
fundamental
strings given in (\ref{eq:generalised}).
The last solution to the best of our knowledge is
new. The
proof that it is free of anomaly related $\alpha'$ corrections is
provided by
the corresponding properties of its dual partner, supersymmetric
string
waves, and by special properties of duality symmetry to preserve the
condition of the embedding of the spin connection in a subgroup of a
gauge group.  Our generalized  fundamental strings are different from
the
charged
heterotic string solution  obtained by Sen \cite{Se2} by twisting the
uncharged
string solution of \cite{Da1}. For instance, in Sen's solution the
charge-dependent
terms in the metric occur in the $du^2 + dv^2$ sector, whereas in our
case the $A_i$-dependent terms occur in the $dudx^i$ sector.
Another difference is that
in our solution the antisymmetric tensor contains $A_i$-dependent
terms, whereas in Sen's solution the antisymmetric tensor has
no charge-dependent terms.
Sen's solution is known to be
supersymmetric in the zero slope limit \cite{Se2,Wa1}.
However, no information is
available
about the $\alpha'$ corrections to this solution.

The advantage of
our method
of using sigma-model duality  is in the fact that the structure of
$\alpha'$ corrections is under control for the
dual solution if it was under control for the original solution.
In this way the nice properties of the plane waves are
carried over to the string-type solutions via sigma model duality,
acting
as a
symmetry of the target space action.

\section*{Acknowledgements}
We are grateful to G. Horowitz for a useful discussion.
The work of E.B.and  R.K. was partially supported by a NATO
Collaborative Research Grant. The work of E.B. has been made
possible by a
fellowship of the Royal Netherlands Academy of Arts and Sciences
(KNAW). E.B.
would like to thank the Physics Department of Stanford University for
its
hospitality. The work  of  R.K.  was supported in part by NSF grant
PHY-8612280.

\newpage

\appendix
\section {Notation and conventions}

We use a metric with mostly minus signature. Our conventions for the
Riemann tensor  and the spin connection are
given in an appendix of \cite{Be1}.
We often use a light-cone basis for the ten-dimensional coordinates
$x^\mu$:
\begin{equation}
x^\mu = (u,v,x^i),\ i=1,\dots ,8, \ \ \  u={1\over \sqrt
2}(t+x),\ v
={1\over \sqrt 2}(t-x)\ ,
\end{equation}
where $t\equiv x^0$ and $x\equiv x^9$.
In this paper, all the indices are raised and lowered with the full
ten-dimensional metric
$g_{\mu\nu}$. In the case in which the metric corresponds to the
SSW solution  given in eq.
(\ref{eq:SSW}), the
relation between upper and lower indices is
\begin{eqnarray}
\xi^{u} & = & \xi_{v}\ \ , \\
\xi^{v} & = & \xi_{u}-(2A_{u}+\sum_{i=1}^{8}A_{i}^{2})\xi_{v}+
\sum_{i=1}^{8}A_{i}\xi_{i}\ \ ,\\
\xi^{i} & = &A_{i}\xi_{v}-\xi_{i}\ \ .
\end{eqnarray}
The constant Killing vectors $k^\mu$ and $l^\mu$ are given in the
light-cone basis by:
\begin{eqnarray}
k^\mu &=& {1\over \sqrt 2} (1,-1,0,\dots,0) \ ,\nonumber\\
l^\mu &=& (0,1,\dots,0)\ .
\end{eqnarray}
The expressions for these Killing vectors with down indices
depend on the metric we are using.  For the SSW metric (\ref{eq:SSW})
they
are given by:
\begin{eqnarray}
k_\mu &=& {1\over 2}\sqrt 2 (2A_u-1, 1, A_i)\ ,\nonumber\\
l_\mu &=& (1,0,\dots , 0)\  .
\end{eqnarray}
The inner products between $k$ and $l$ for the SSW metric
are given by
\begin{equation}
l^2 = 0\ ,\hskip 1truecm k^2=A_u-1\ ,\hskip 1truecm k \cdot l =
{1\over 2}
\sqrt 2\ .
\end{equation}

\section{Proof of Supersymmetry of the Generalized FS
Solution}

In this appendix we will show that the generalized FS solution
is supersymmetric.
Since the
supersymmetry transformation rules involve fermions it is necessary
to reformulate the SSW Ansatz for the metric in terms of zehnbein
fields:
\begin{equation}
e_\mu{}^a = \delta_\mu{}^a + A_\mu l^a\ .
\end{equation}
The unbroken supersymmetries of the SSW solution are given by
\begin{equation}
\label{eq:g1}
l^\mu\gamma_\mu\epsilon_0 = 0 \hskip .5truecm {\rm or}
\hskip .5truecm (\gamma^0+\gamma^9)\epsilon_0 = 0\ ,
\end{equation}
where $\epsilon_0$ is a constant ten-dimensional spinor.
It is instructive to see how the sigma model duality transformation
leads to unbroken supersymmetries for the generalized FS solution
as well.

In order to investigate the existence of unbroken supersymmetries for
our purely bosonic
solutions we only need to consider the bosonic terms in the
supersymmetry
transformation rules of the fermions.  They are given by
\begin{eqnarray}
\delta \psi_\mu &=& \bigl (\partial_\mu - {1\over 4}
  \Omega_{\mu +}{}^{ab}\gamma_{ab}\bigr)
\epsilon   \ ,
   \label{eq:susy1}\\
\delta\lambda &=& \bigl (\gamma^\mu\partial_\mu\phi + {1\over 4}
                               H_{\mu\nu\rho}
\gamma^{\mu\nu\rho}\bigr)\epsilon \ ,
                              \label{eq:susy2}\\
\delta\chi &=& - {1\over 4} F_{\mu\nu}\gamma^{\mu\nu}\epsilon\ .
\end{eqnarray}
We first consider the $\lambda$ transformation rule. Requiring that
$\delta\lambda = 0$ leads to the equation
\begin{equation}
\label{eq:dl=0}
\gamma^i(\partial_i\phi)\epsilon + {3\over
2}H_{iuv}\gamma^{iuv}\epsilon
+ {3\over 4}H_{iju}\gamma^{iju}\epsilon
 =  0\ ,
\end{equation}
where it is understood that in this equation the generalized FS
solution is
substituted. To investigate this equation we need the form of the
(inverse)
zehnbein fields of the generalized FS solution. They are given by:
\begin{eqnarray}
e_a^u &=& \delta^v_a \ ,\nonumber\\
e^v_a &=&  (1-A_u)\delta^u_a - A_i\delta^i_a\ .\\
e^i_a &=& \delta^i_a\nonumber\ .
\end{eqnarray}
It is now not too difficult to show that
the first two and the last terms in (\ref{eq:dl=0}) vanish separately
provided that the supersymmetry parameter
$\epsilon$
satisfies\footnote{Note that both (\ref{eq:g1}) and
(\ref{eq:condsusy})
can be written as $\gamma^u\epsilon =0$, using curved indices.}
\begin{equation}
(\gamma^0 - \gamma^9)\epsilon = 0\ .
\label{eq:condsusy}
\end{equation}
We next consider the gravitino transformation rule.  Instead of
substituting the
generalized FS solution into the equation
$\delta\psi_\mu=0$ it is easier, and equivalent,  to use the SSW
solution and
to
 require that after a duality transformation $\delta\psi_\mu=0$.
This leads to the equation
\begin{equation}
\delta\psi_\mu = \bigl (\partial_\mu - {1\over 4}
\tilde {\Omega}_{\mu+}{}^{ab}\gamma_{ab}\bigr )\epsilon = 0\ .
\end{equation}
Applying the duality rotation of the torsionful spin connection given
in
(\ref{eq:dualtorsion}) we find
\begin{equation}
\partial_\mu\epsilon -{1\over
4}\Omega_{\mu+}{}^{ab}\gamma_{ab}\epsilon
+ {1\over k^2}k^\lambda k^a\Omega_{\lambda -,\mu}{}^b\gamma_{ab}
\epsilon = 0
\ ,
\end{equation}
where it is now understood that the SSW solution is substituted.
We  next substitute the expression for the torsionful spin
connections
corresponding to the SSW solution \cite{Be1}:
\begin{equation}
\Omega_{\mu+}{}^{ab} = -2 l^{[a}{\cal A}^{b]}{}_\mu\ ,\hskip 1truecm
\Omega_{\mu-}{}^{ab} =  {\cal A}^{ab}l_\mu\ ,
\end{equation}
where ${\cal A}_{\mu\nu} = \partial_\mu A_\nu - \partial_\nu A_\mu$.
We thus find that the following equation must be satisfied:
\begin{equation}
\partial_\mu\epsilon + {1\over 2} l_\rho {\cal A}_{\sigma\mu}
\gamma^{\rho\sigma}\epsilon + {k \cdot l\over k^2} k_\rho
{\cal A}_{\mu\sigma}\gamma^{\rho\sigma}\epsilon = 0\ .
\end{equation}
The $\mu=v$ component of this equation is automatically satisfied.
Using the condition  (\ref{eq:condsusy}) we find that the $\mu = u$
component is satisfied as well. Finally, for $\mu = i$ we find that
the equation is satisfied provided that
\begin{equation}
\epsilon = (A_u-1)^{-1/2}\epsilon_0
\label{eq:condsusy2}
\end{equation}
where $\epsilon_0$ is constant.

Finally, by using the condition (\ref{eq:condsusy}) in the form
$\gamma^u\epsilon=0$ and the fact that the gauge field corresponding
to the
generalized FS solution only has a nonvanishing $u$-component, it is
not too difficult to show that also $\delta\chi = 0$.

We thus conclude that the generalized FS solution has 8
 unbroken supersymmetries given by eqs.~(\ref{eq:condsusy})
and (\ref{eq:condsusy2}).
 We should stress that it is not obvious that duality
transformations {\sl always} preserve the supersymmetry of a given
solution.
Note that the unbroken supersymmetries before and after the
duality transformation differ from each other.

\section{Proof of the space-time supersymmetry of the
K-T Solution}
The solution of \cite{KT} contains a flat metric describing
a four-dimensional spacetime with coordinates
$x^\mu=\{u,v,x^1,x^2\}$,
a 2-form field
\begin{equation}
B= - 2 f(u) dx^1\wedge dx^2\ ,
\end{equation}
and a $u$-dependent dilaton $\phi(u)$.
The only nonzero component of $H$ is given by
 $H_{u12}=-1/3 \partial_u f$.

The supersymmetry transformation rules for the dilatino
\begin{equation}
\delta\lambda = \bigl (\gamma^\mu\partial_\mu\phi + {1\over 4}
                               H_{\mu\nu\rho}
\gamma^{\mu\nu\rho}\bigr)\epsilon = 0
\end{equation}
have a non-trivial solution under the condition that
\begin{equation}
\gamma^u \epsilon = 0\ .
\end{equation}
The supersymmetry transformation rules for the gravitino have a $v$-
component
\begin{equation}
\delta \psi_v = \bigl (\partial_v - {1\over 4}
  \Omega_{v +}{}^{ab}\gamma_{ab}\bigr)
\epsilon   = 0\ .
\end{equation}
This equation is satisfied under the condition that $\epsilon$ is
$v$-independent, since $\Omega_{v +}{}^{ab}=0$.
The $i$ component of this equation is solved if $\epsilon$ is
$x^i$-independent and $\gamma^u \epsilon = 0$:
 \begin{equation}
\delta \psi_i = \bigl (\partial_i - {1\over 4}
  \Omega_{i +}{}^{ab}\gamma_{ab}\bigr)
\epsilon = 0  \ .
\end{equation}
The $u$-component of this equation
 \begin{equation}
\delta \psi_u = \bigl (\partial_u + {3\over 4}
  H_{u}{}^{12}\gamma_{12}\bigr)
\epsilon  = 0 \ ,
\end{equation}
is satisfied if the spinor of unbroken
supersymmetry $\epsilon$ has a specific dependence on $u$-coordinate
of
the form
\begin{equation}
\epsilon (u) =  e^{{1\over 4}f(u) \gamma^1 \gamma^2 } \epsilon _0\ .
\end{equation}

After a duality transformation in the $x^1$-direction,
the dual solution is given by a
zero axion field and the following non-flat metric:
\begin{equation}
ds^2 = 2dudv + 2f(u)dx^1dx^2 - dx^1dx^1 - (1+f^2)dx^2dx^2\ .
\end{equation}
The dilaton field remains unchanged: $\tilde\phi = \phi$.
The supersymmetry of the dual solution is proven as follows. The
equation $\delta\lambda=0$ again leads to the condition that
$\gamma^u\epsilon=0$. To investigate the supersymmetry transformation
rule of the gravitino it is convenient to first consider the
dual of $\Omega_+$:
\begin{equation}
{\tilde \Omega}_{c+}{}^{ab} = -{3\over 2} (H_c{}^{ab} - 4 k^{[a}
H_{1c}{}^{b]})\ .
\end{equation}
Here $k^\mu$ is the vector $k^\mu = (0,0,1,0)$.
Using this expression it follows that the equation $\delta \psi_v =
0$
is satisfied under the condition that $\epsilon$ is $v$-independent.
Furthermore, using $\gamma^u\epsilon = 0$, it follows that
$\delta\psi_i=0$ if $\epsilon$ is $x^i$-independent.
Finally, the equation $\delta\psi_u=0$ is satisfied if
 \begin{equation}
\delta \psi_u = \bigl (\partial_u - {3\over 4}
  H_{u}{}^{12}\gamma_{12}
\bigr)
\epsilon  = 0 \ ,
\end{equation}
or
\begin{equation}
\epsilon (u) =  e^{-{1\over 4} f(u) \gamma^1 \gamma^2 } \epsilon _0\
{}.
\end{equation}

Finally, one can show that both before and after the duality
transformation all the $T$-tensors and Lorentz Chern-Simons terms
vanish.
We therefore conclude that both the Klim\v c\'\i k-Tseytlin solution
as well as
its
dual are supersymmetric to all orders in $\alpha'$.

\pagebreak


\begin{thebibliography}{100}

\bibitem{Ca1}C.G.~Callan, Jr., J.A.~Harvey and A.~Strominger,
{\sl Supersymmetric String Solitons}, proceedings of the 1991 Trieste
Spring School on String Theory and Quantum Gravity
 (World Scientific, 1992).

\bibitem{Ho2}G.T.~Horowitz, {\sl The Dark Side of String Theory:
Black Holes and Black Strings}, to appear in the proceedings of the
1992 Trieste Spring School on String Theory and
Quantum Gravity.

\bibitem{Se1} A.~Sen, {\sl Black Holes and Solitons in String
Theory}, to appear in the proceedings of the 1992 ICTP
 Summer Workshop, Trieste, July 2-3.

\bibitem{Be2} E. Bergshoeff and M. de Roo, Nucl. Phys.
{\bf B328}, 439 (1989).

\bibitem{St1}A.~Strominger, Nucl.~Phys.~{\bf B343}, 167 (1990);
C.G.~Callan, Jr., J.A.~Harvey and A.~Strominger,
 Nucl.~Phys.~{\bf B359}, 611 (1991).

\bibitem{BUS} T. Buscher, Phys. Lett. {\bf  159B}, 127 (1985);
 {\it  ibid} {\bf
194B}, 59 (1987); {\it ibid} {\bf  201B}, 466 (1988).

\bibitem{Ve1} G.~Veneziano, Phys.~Lett.~{\bf B265}, 287 (1991);
K.~Meissner and G.~Veneziano, Phys.~Lett.~{\bf B267},  33 (1991);
M.~Gasperini and G.~Veneziano, Phys.~Lett.~{\bf B277},  256 (1992).

\bibitem{Se2}A.~Sen, Phys.~Lett.~{\bf B271},  295 (1991); {\it ibid.}
Phys.~Lett.~{\bf B274},  34 (1992); {\it ibid.}
Nucl.~Phys.~{\bf B388},  457 (1992);
S.~Hassan and A.~Sen,
Nucl.~Phys.~{\bf B375},  103 (1992).

\bibitem{KT} C.~Klim\v c\' \i k and A.A.~Tseytlin, {\sl Duality
invariant class of exact string backgrounds}, preprint
CERN-TH.7069/93.

\bibitem{KG} E.~Kiritsis, Nucl.~Phys.~B405 (1993) 109; A.~Kiritsis
and A.~Giveon, preprint hepth/9303016, Nucl.~Phys.~B in press;
{\it ibid.} preprint hepth/9309064.

\bibitem{GE} D.~Gershon, preprint hepth/9311122.


\bibitem{Be1} E. Bergshoeff, R. Kallosh and T. Ort\'\i n, Phys. Rev.
{\bf D47},  5444 (1993).

\bibitem{Ho4} J.H.~Horne, G.T.~Horowitz and A.R.~Steif,
Phys.~Rev.~Lett.~{\bf 68},  568 (1992).

\bibitem{Da1} A.~Dabholkar, G.~Gibbons, J.~Harvey and F.~Ruiz,
Nucl.~Phys.~{\bf B340},  33 (1990).

\bibitem{Gu1}
R. G\"{u}ven, Phys. Lett. {\bf 191B},  265 (1987).

\bibitem{Am1}
D. Amati and  C. Klim\v{c}\' \i k, Phys. Lett {\bf 219B},  443
(1988);
D.
Amati, M.
Ciafaloni and Veneziano, Nucl. Phys.  {\bf B347},  550 (1990).

\bibitem{Ho1}
G.T. Horowitz and A. R. Steif, Phys. Rev. Lett. {\bf 64},  260
(1990); {\sl ibid.} Phys.~Rev.~{\bf D42},  1950 (1990).

\bibitem{HW} C. M. Hull and E. Witten, Phys. Lett. {\bf  160B}, 398
(1985).

\bibitem{RV} M. Ro\v cek and E. Verlinde, Nucl. Phys. {\bf  B373},
630
(1992), and references therein.

\bibitem{Ch1}
A.H. Chamseddine, Nucl. Phys. {\bf B185},  403 (1981);
E. Bergshoeff, M. de Roo, B. de Wit and P. van Nieuwenhuizen,
Nucl. Phys. {\bf  B195},  97 (1982);
G.F. Chapline and N.S. Manton, Phys. Lett. {\bf  120B},  105 (1983).

\bibitem{Ru1} R.E.~Rudd, Nucl.~Phys.~{\bf B352},  489 (1991).

\bibitem{Ho3} G.T.~Horowitz, in {\it Proceedings of Strings '90},
College
Station (World Scientific, 1991).

\bibitem{Br1} H.W.~Brinkmann, Proc.~Natl.~Acad.~Sci.~U.S.\,~{\bf 9},
1
(1923).

\bibitem{deRoo} M.~de Roo, M.~Suelmann and A.~Wiedemann,
Nucl.~Phys.~{\bf
405},  326 (1993); M.~Suelmann, {\sl Effective Actions for Heterotic
String Theory},
to appear in the proceedings of the {Journ\'ees Relativistes '93}.

\bibitem{Wa1} D.~Waldram, Phys.~Rev.~{\bf D47},  2528 (1993).

\end{thebibliography}
\end{document}